\documentclass[a4paper,11pt]{article}
\pdfoutput=1 
\usepackage{jinstpub}
\title{Search for light dark matter with NEWS-G}

\author{Konstantinos Nikolopoulos on behalf of NEWS-G collaboration}
\affiliation{School of Physics and Astronomy, University of Birmingham,\\Birmingham, B15 2TT, United Kingdom}
\emailAdd{k.nikolopoulos@bham.ac.uk}

\abstract{The NEWS-G collaboration is searching for light dark matter
  candidates using a spherical proportional counter. Light gases, such
  as hydrogen, helium, and neon, are used as targets, providing access
  in the 0.1--10~\gev\ mass range. First results obtained with SEDINE,
  a 60 cm in diameter detector, in the Underground Laboratory of
  Modane yielded a 90\% confidence level upper limit of
  $4.4\cdot 10^{37}$~cm$^{2}$ on the nucleon-dark matter interaction
  cross-section for a candidate with 0.5~\gev\ mass. Recent
  developments in several aspects of the detector instrumentation are
  presented, along with the construction of a new, 140 cm in diameter,
  detector with new compact shielding.}

\keywords{Dark Matter detectors (WIMPs, axions, etc.); Gaseous detectors; Materials for gaseous detectors; Gas systems and purification}

\proceeding{15$^{\text{th}}$ Topical Seminar on Innovative Particle and Radiation Detectors (IPRD19)\\
14-17 October 2019\\
Siena, Italy}

\usepackage{atlasphysics}
\usepackage{wrapfig}
\usepackage{paralist}
\usepackage{subfigure}
\usepackage{enumitem}
\usepackage{textgreek}

\usepackage{siunitx}
\DeclareSIUnit\parsec{pc}
\DeclareSIUnit\lightyear{ly}
\DeclareSIUnit\HOUR{hour}
\DeclareSIUnit\DAY{day}
\DeclareSIUnit\YEAR{year}

\usepackage{amssymb,wasysym}
\usepackage{blindtext}
\usepackage{scrextend}
\addtokomafont{labelinglabel}{\sffamily}

\usepackage{tikz}
\begin{document}
\maketitle
\flushbottom

\section{Introduction}
It is established from a variety of astrophysical
observations~\cite{Bertone:2010zza,1538-4357-648-2-L109} and precise
measurements of the Cosmological Microwave
Background~\cite{Planck2015} that approximately 84.5\% of the matter
content of our Universe consists of non-baryonic cold Dark Matter
(DM).  Although the nature of DM is currently unknown, many theories
beyond the Standard Model (SM) predict massive neutral particles,
thermally produced in the early Universe, that could account for the
observed DM relic density. A generic class of well motivated DM
candidates is known as Weakly Interacting Massive Particles
(WIMPs)~\cite{Feng:2010gw}. The WIMP hypothesis favours DM masses in
the 10--1000~\gev\ range~\cite{Jungman:1995df}. However, the lack of
evidence for supersymmetry at the LHC~\cite{Buchmueller:2012hv} and of
convincing evidence from direct and indirect detection experiments
motivates the investigation of models with lighter DM candidates and,
potentially, more complex couplings --- e.g. hidden
sectors~\cite{Essig:2013lka,Profumo:2015oya}, asymmetric dark
matter~\cite{Petraki:2013wwa,Zurek:2013wia}, and more generic
descriptions through effective theory~\cite{Schneck:2015eqa}.

Direct detection experiments aim to detect incoming DM particles from
the Milky Way halo via their coherent elastic scattering off a target
nuclei. A key element towards improved sensitivity to low mass
candidates is the possibility to operate with low energy threshold,
dictated by the expected nuclear recoil energy spectrum, which is
concentrated to ever lower energies as the DM candidate mass
decreases. A further experimental challenge arises from the fact that
recoil ions only partially dissipate their kinetic energy as
ionisation. The fraction of ion kinetic energy released in the detector
medium through ionisation, the ionisation quenching factor, depends on the ion atomic
number and kinetic energy. Sensitivity to low mass DM candidates is
further hindered by the rapid decrease of this factor with decreasing kinetic
energy, for energies below a few keV~\cite{Lindhard1963}.

\section{The spherical proportional counter}
\begin{wrapfigure}{L}{0.45\linewidth}
\centering
\vspace{-0.6cm}
\includegraphics[width=0.95\linewidth]{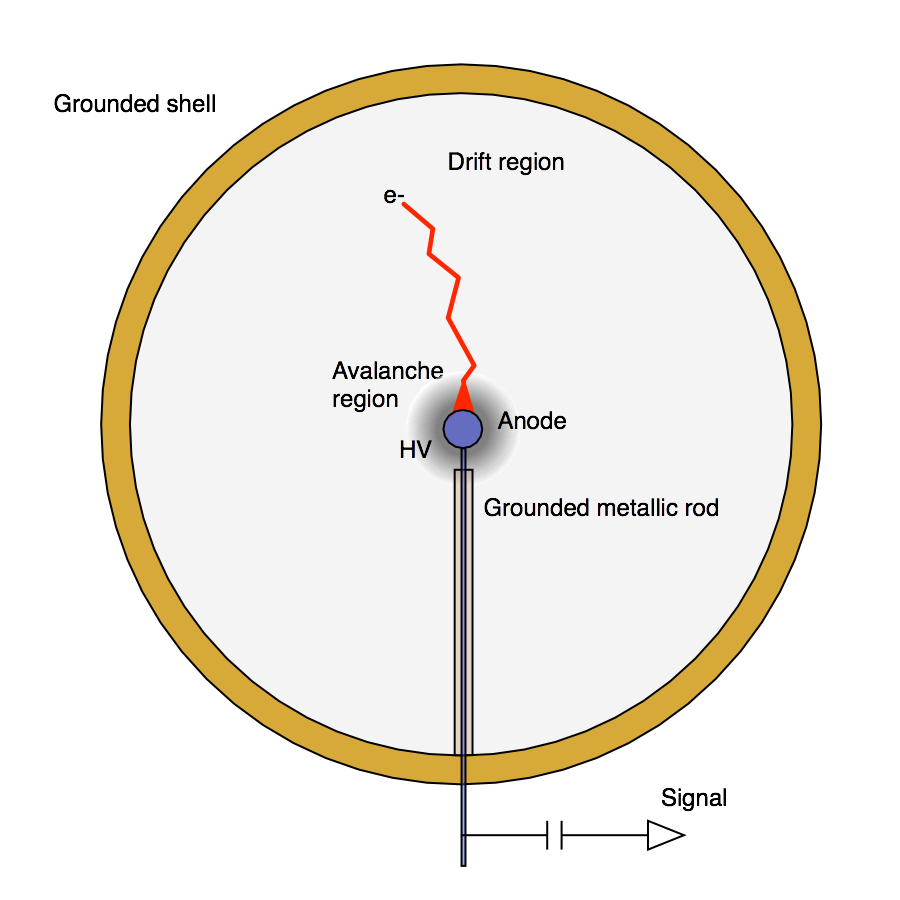}
\vspace{-0.5cm}
\caption{Spherical proportional counter design and principle of operation~\cite{Savvidis:2016wei}.\label{fig:spc_draw}}
\end{wrapfigure}

The spherical proportional counter, presented in figure~\ref{fig:spc_draw},
is a novel gaseous detector~\cite{Giomataris:2008ap, Gerbier:2014jwa,
  Savvidis:2016wei}. It consists of a grounded spherical shell which
acts as the cathode and a small spherical anode, the sensor, supported
at the centre by a grounded metallic rod, to which the high voltage is
applied and from which the signal is read-out. In the ideal case the
electric field has an $1/r^2$ dependence on the radial distance from
the detector centre. This dependence naturally divides the detector
into the drift region, where under the influence of the electric
field the electrons drift towards the anode, and the amplification
region, where charge multiplication occurs.

The spherical geometry provides several advantages for building large
volume detectors. The sphere has the lowest surface-to-volume ratio
and is well suited for high pressure operation. Furthermore, light
elements exhibit favourable ionisation quenching factor. Overall, the
spherical proportional counter exhibits the following key features:
\begin{inparaenum}[a)]
\item {very low energy thresholds}, down to single ionisation electron detection, thanks to small sensor capacitance and high gain operation;
\item {small number of read-out channels};
\item {background rejection and fiducialisation} through pulse shape analysis;
\item {simple and robust construction with radiopure materials};
\item {variety of light target gases},
allowing optimisation of momentum transfer for light particles; and
\item {possibility to vary the operational pressure and high voltage}, providing additional handles to
disentangle potential signals from unknown backgrounds.
\end{inparaenum}

\section{First NEWS-G results on the search for light DM}
\begin{figure}
\centering
\subfigure[\label{fig:sedine1}]{\includegraphics[width=0.25\linewidth]{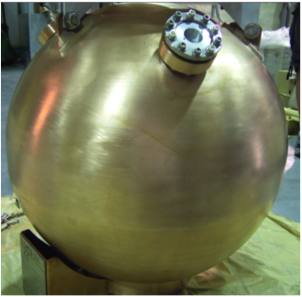}}
\subfigure[\label{fig:sedine2}]{\includegraphics[width=0.15\linewidth]{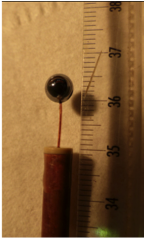}}
\subfigure[\label{fig:sedine3}]{\includegraphics[width=0.36\linewidth]{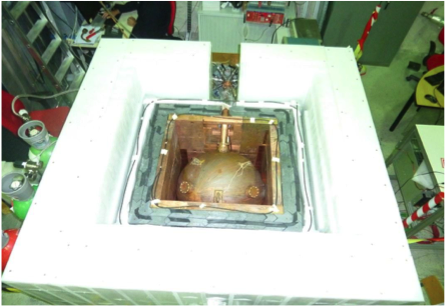}}
\caption[]{\subref{fig:sedine1} SEDINE, the $\varnothing$60~cm spherical proportional counter operating at LSM; \subref{fig:sedine2} the $\varnothing$6.3~mm silicon sensor installed in SEDINE, with a $\varnothing$380~$\mu$m diameter insulated HV wire routed through a grounded copper rod; and \subref{fig:sedine3} the cubic shielding of SEDINE.\label{fig:spc}}
\end{figure}

The $\varnothing$60~cm spherical proportional counter SEDINE, shown in
figure~\ref{fig:sedine1}, is installed
at the Laboratoire Souterrain de Modane (LSM), which has an overburden
of 4800~m water equivalent.  The detector is constructed using pure (NOSV) copper,
chemically cleaned to remove radon deposits, and a $\varnothing$6.3~mm
spherical anode made of silicon, shown in figure~\ref{fig:sedine2}. In
figure~\ref{fig:sedine3}, the shielding of SEDINE is shown, which --- moving outwards --- comprises 8~cm of copper, 15~cm of lead, and 30~cm of polyethylene. First
results were obtained using a ${\rm Ne}$/${\rm CH_4}$ (99.3\%/0.7\%)
gas mixture at 3.1~bar pressure, with a total target mass exposure of $9.6\;{\rm kg}\cdot{\rm days}$.

\begin{figure}[h]
\centering
 \includegraphics[width=0.6\textwidth]{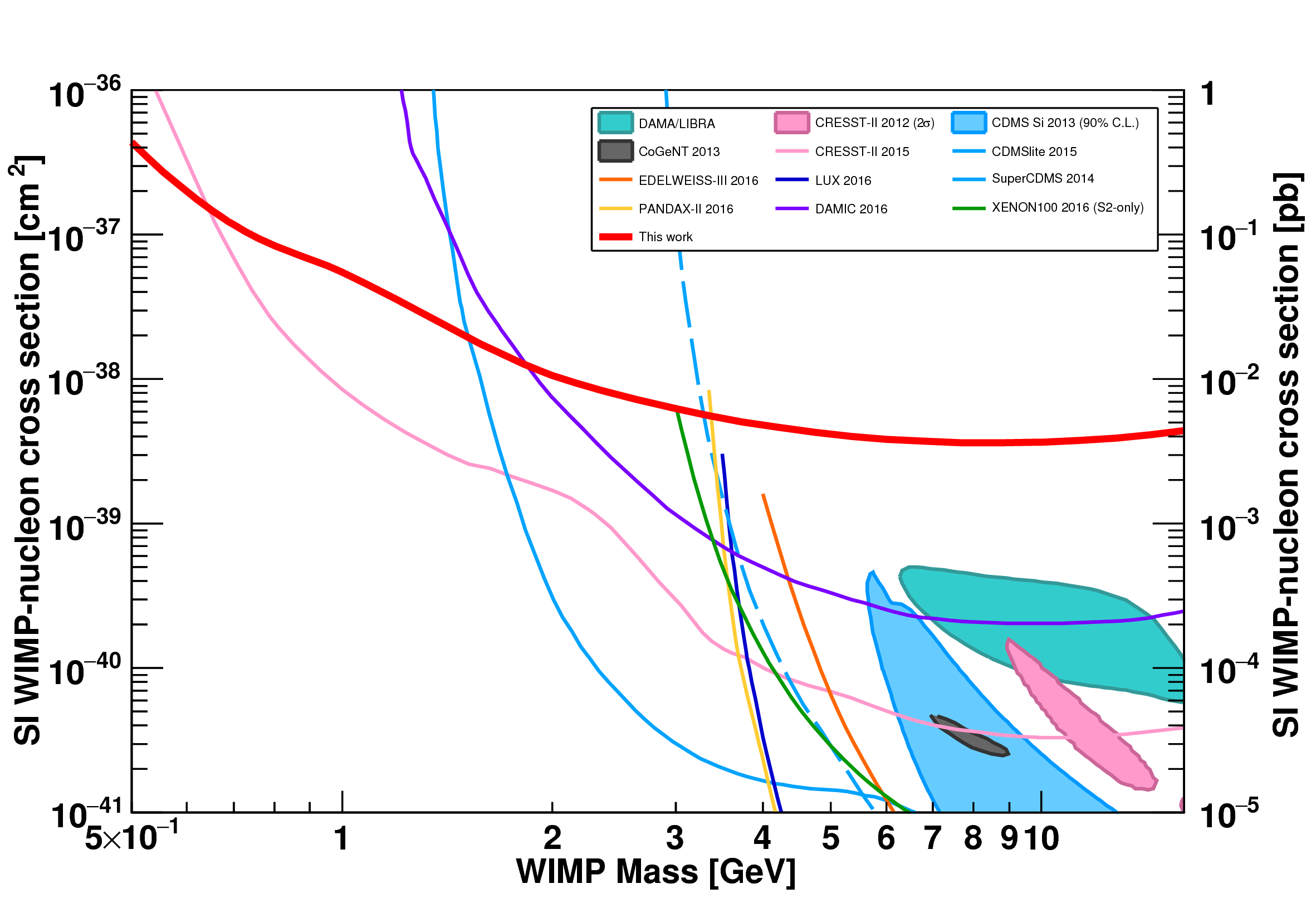}
\caption{NEWS-G constraints in the spin-independent WIMP-nucleon cross
  section as a function of the WIMP mass plane (solid red line)~\cite{Arnaud:2017bjh}.\label{fig:news}}
\end{figure}

The 90\% confidence level (CL) upper limit on the spin-independent
DM-nucleon scattering cross section, derived considering all observed events as candidates, is presented in figure~\ref{fig:news} as a
solid red line~\cite{Arnaud:2017bjh}. New constraints for masses below 0.6~GeV are set. The recoil energy
spectrum used to derive the sensitivity to light DM is based on
standard assumptions of the DM-halo model.\footnote{DM density
  $\rho_{DM} = 0.3\;{\rm GeV}/{\rm cm}^{3}$, galactic escape velocity
  $v_{\rm esc} = 544\;{\rm km/s}$, asymptotic circular velocity $v_{0}
  = 220\;{\rm km/s}$.}

\section{Developments in sensor design}
The design of the sensor and its support structure are crucial for the
detector performance, as they directly affect the electric field
strength and uniformity. In the following, recent developments on this
topic are discussed.

\subsection{Single-anode sensors}
The anode support structure, in the simplest case consisting of a
grounded rod, affects the electric field homogeneity and, thus, the
homogeneity of the detector response. This is improved with the
development of support structures with correction electrodes, more
recently through the development of the resistive glass
electrode~\cite{Katsioulas:2018pyh}.

\begin{wrapfigure}{L}{0.35\linewidth}
\centering
\vspace{-0.2cm}
\includegraphics[width=0.27\textwidth]{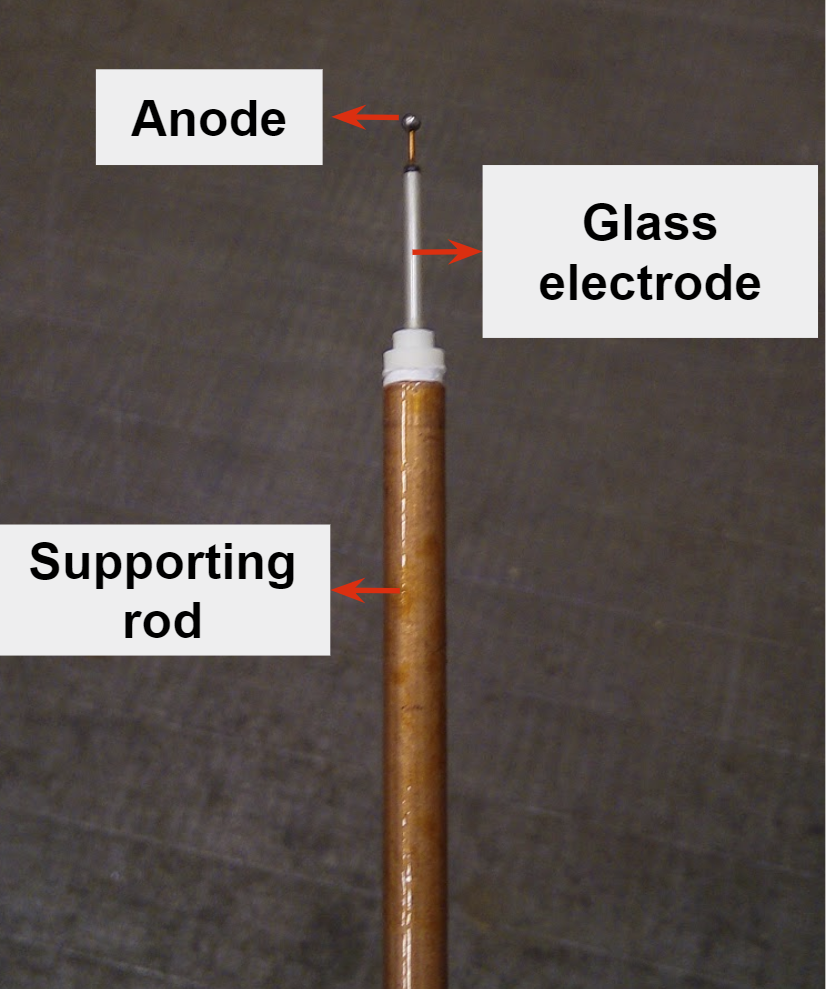}
\vspace{-0.2cm}
\caption[]{Module with a cylindrical glass correction electrode~\cite{Katsioulas:2018pyh}.\label{fig:sensor_real}}
\end{wrapfigure}

In figure~\ref{fig:sensor_real}, the constructed module is presented,
composed of a $\varnothing 2\;{\rm mm}$ anode made of stainless
steel. The glass tube has a length of $20\;{\rm mm}$ and the distance
between the tube and the anode surface is $3\;{\rm mm}$. The module
was tested in a $\varnothing 30~\si{\cm}$ spherical, stainless steel
vessel, supported by a copper rod with a $4~\si{\mm}$ ($6~\si{\mm}$)
inner (outer) diameter.

\begin{figure}[htbp]
\centering 
\subfigure[\label{fig:homogeneityFigures}]{\includegraphics[width=0.40\linewidth]{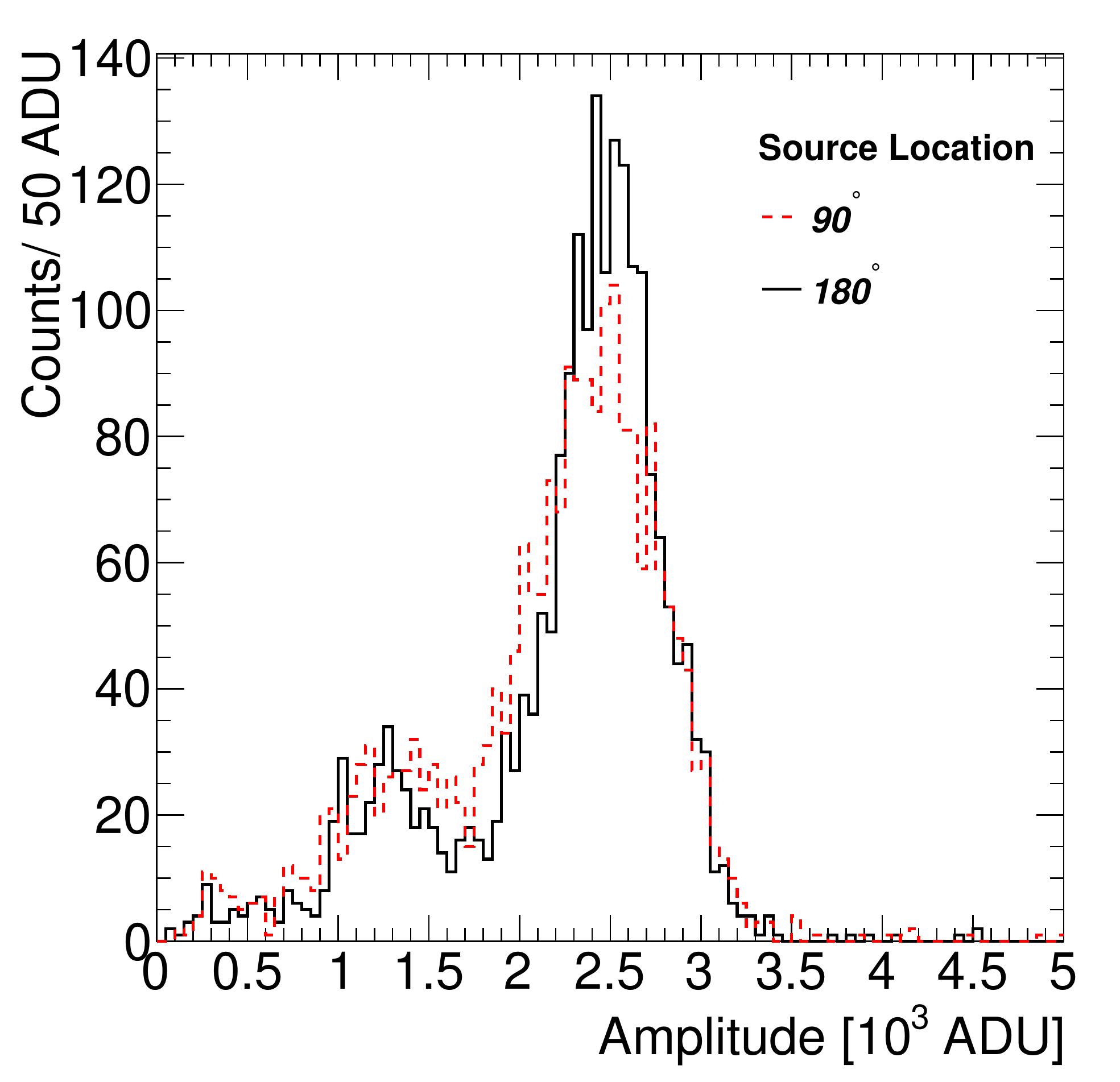}}
\subfigure[\label{fig:stability}]{\includegraphics[width=0.4\linewidth]{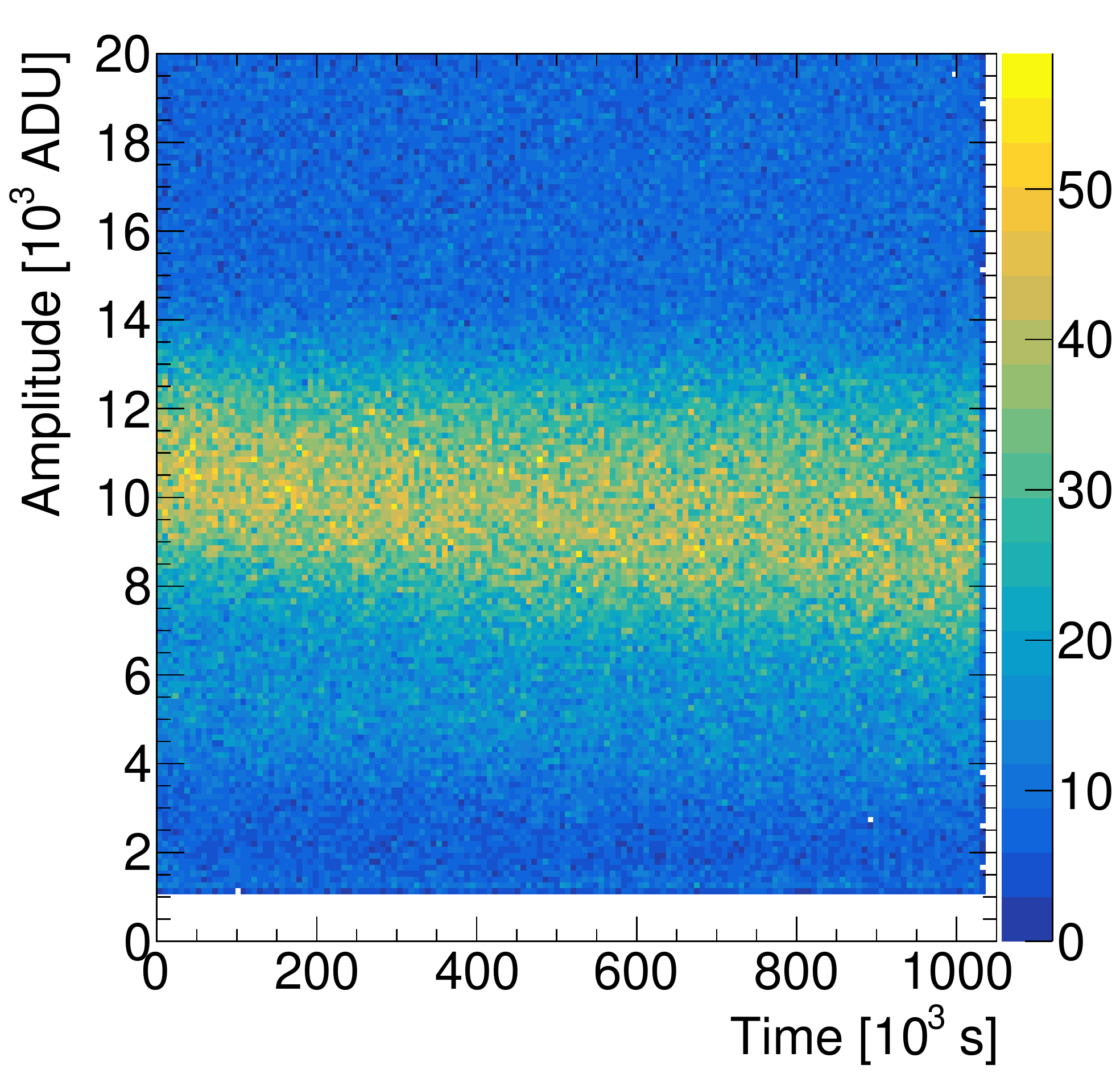}}
\caption{\subref{fig:homogeneityFigures} Pulse amplitude distribution for
  $5.9~\si{keV}$ X-rays from an $^{55}$Fe source located inside the
  detector placed at zenith angle of $\ang{90}$ and $\ang{180}$,
  relative to the grounded rod. The detector is filled with
  He:Ar:CH$_{4}$ (92\%:5\%:3\%) at $1~\si{bar}$.
  \subref{fig:stability} Pulse amplitude as a function of time using a
  detector filled with $2~\si{bar}$ of He:Ar:CH$_{4}$ (87\%:10\%:3\%).
  Both figures from ref.~\cite{Katsioulas:2018pyh}. }
\end{figure}

The electric field homogeneity was investigated using an $^{55}{\rm
  Fe}$ source placed inside the detector. The position of this
collimated source could be modified during detector operation.  Data
were collected with the source located at $\ang{90}$ and $\ang{180}$
to the grounded rod and the distribution of signal
amplitude is shown in figure~\ref{fig:homogeneityFigures}, demonstrating
similar response in both cases and, thus, improved uniformity.

The detector operation stability was tested using He:Ar:CH$_{4}$
(87\%:10\%:3\%) at $2~\si{bar}$, introduced using a filter to remove
oxygen and water vapour.  The $6.4~\si{keV}$ X-ray fluorescence of the
$^{55}$Fe K-line, induced by environmental $\gamma$-rays and cosmic
muons, was used to monitor the gas gain as a function of time, as
shown in figure~\ref{fig:stability}, for 12~days of continuous
data-taking. The detector was stable throughout the entire period,
with no spark-induced gain variations. The observed decrease
of the pulse amplitude in the timescale of days is the result of the gradual introduction of contaminants in the gas volume.

\subsection{Multi-anode sensors}
One of the challenges towards development of large size spherical
proportional counters is the inter-dependence of the detector gain
and the electron drift velocity, in particular at large radii, through the anode 
electric field in the single-anode sensor. By increasing the electric field of
the anode the ionisation electrons at large
radii would be efficiently collected, but could lead to breakdown during the avalanche creation,
while an electric field providing an acceptable gas gain could be
inefficient for the effective collection of the charges.

\begin{wrapfigure}{L}{0.40\linewidth}
\centering
\vspace{-0.5cm}
\includegraphics[width=0.25\textwidth]{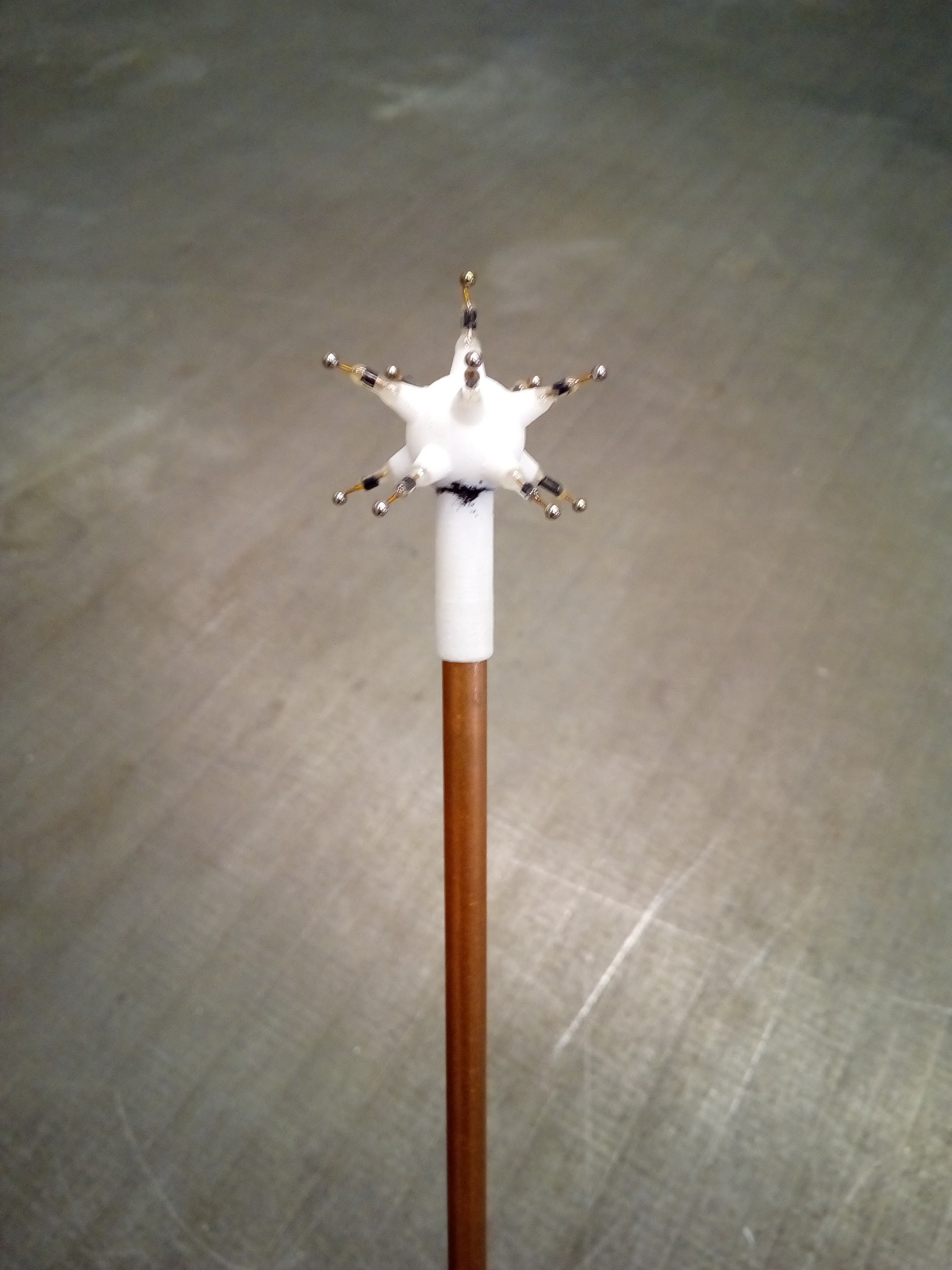}
\vspace{-0.2cm}
\caption{ACHINOS with 11 balls of $\varnothing$2~mm constructed using 3D printing~\cite{Giganon:2017isb}.\label{fig:achinos1}}
\end{wrapfigure}

For this reason, ACHINOS~\cite{Giganon:2017isb} --- a new multi-anode
sensor --- is being developed, composed of multiple anode balls
equidistantly placed on a virtual spherical surface and all biased at
the same potential, as shown in
figure~\ref{fig:achinos1}. Collectivelly, this leads to an increased
electric field strength at large radii, while maintaining the ability
to reach high gain operation provided by the individual anode
diameter.  For example, an ACHINOS sensor with 11 anodes distributed on
a $\varnothing$36~mm sphere produces approximately 9 times larger
electric field at large radii, with respect to a single $\varnothing$2~mm
anode at the same bias voltage~\cite{Giganon:2017isb}. This
development paves the way for large detector operation under high
pressure.

\section{A large size spherical proportional counter with compact shielding}
\begin{figure}[htbp]
\centering
\includegraphics[width=0.49\textwidth]{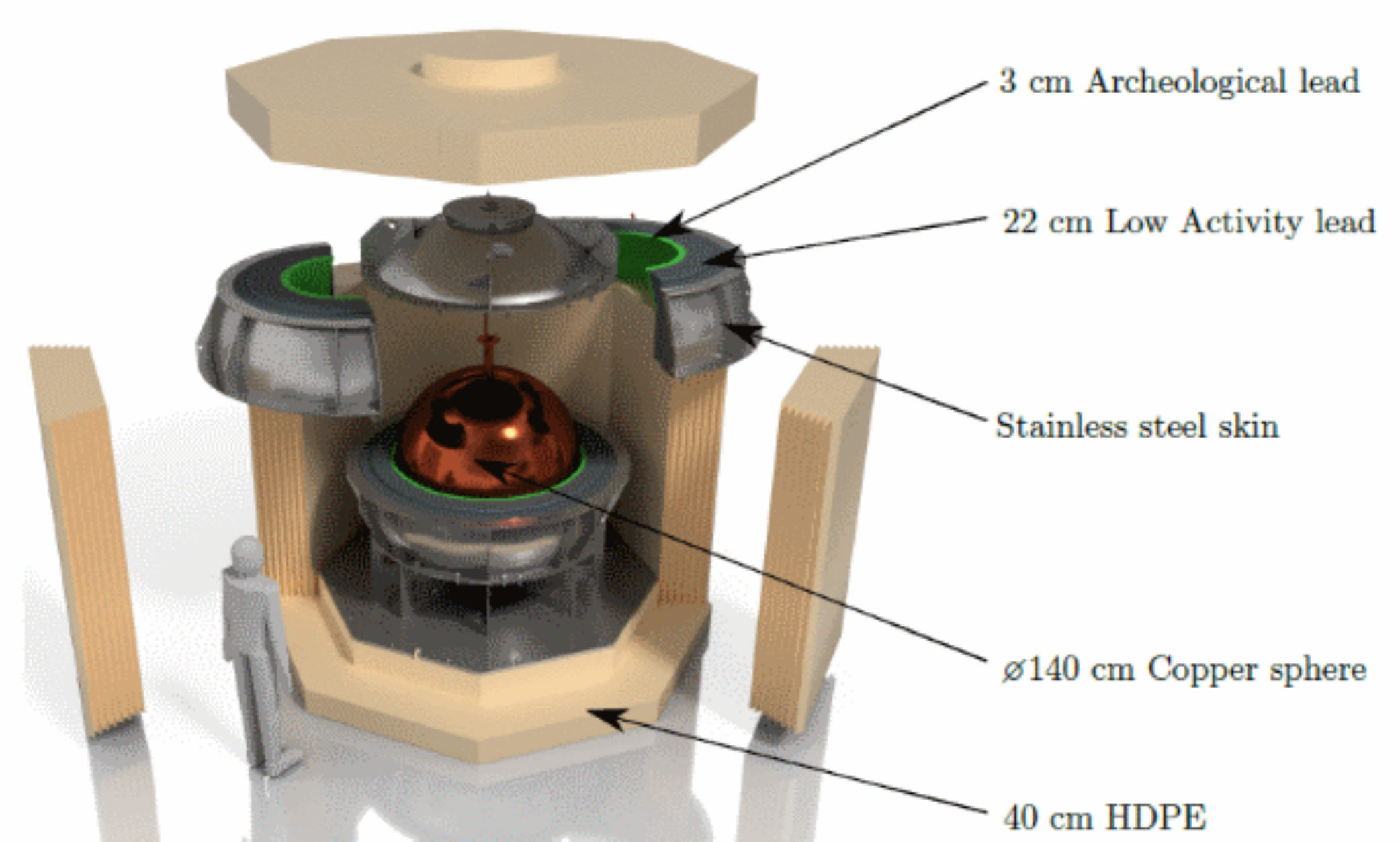}
\caption{Schematic of the $\varnothing$140~cm spherical proportional counter and its shielding.\label{fig:newsgSNO}}
\end{figure}

The next phase of the experiment builds on the experience acquired
from the operation of SEDINE at LSM and consists of a
low-background $\varnothing$140~cm spherical proportional counter at the
centre of a new compact shielding. The spherical proportional counter
comprises two hemispheres made of Aurubis C10100 copper, with a purity
of 99.99\%, that were electron beam welded together.  The shielding
consists of a shell with 3~cm of archaeological lead and 22~cm
low-activity lead, which is placed inside a 40~cm thick polyethylene
shield, as shown in figure~\ref{fig:newsgSNO}. Initial commissioning
of the spherical proportional counter took place at LSM and,
subsequently, the detector was transfered to SNOLAB, with an
overburden of 6000~m water equivalent, for the main physics run. These
improvements are expected to lead to significant background reduction,
relative to that of SEDINE, and will allow sensitivity down to cross
sections of $\mathcal{O}$ $(10^{-41}\;{\rm cm}^{2})$ for masses around
1~\gev. The use of hydrogen and helium-rich targets will enable unprecedented 
experimental sensitivity down to DM candidate masses of 0.1 GeV.

\section{Electroplating for NEWS-G}
Recently, it has been shown that even class 1,\footnote{Classification according to the
American Society for Testing and Materials (ASTM) B170 C10100
standard.} oxygen-free copper contains unacceptably large amounts of
${}^{210}$Po and ${}^{210}$Pb~\cite{Abe:2017jzw}, significantly
deteriorating the experimental sensitivity. Geant4-based
simulations~\cite{Agostinelli:2002hh} for the $\varnothing$140~cm detector suggested that
$^{210}$Pb and $^{210}$Bi decays would be contributing
$4.6\;\si{dru}$,\footnote{$1\;\si{dru} = 1\; \si{count \per\kilo\eV\per
    \kilo\gram\per \DAY}$} below $1\;\si{\kilo\eV}$,
%4.58
approximately an order of magnitude larger than other background
contributions in this energy range.  Thus, it was decided to
electroplate a $500\;\si{\micro\meter}$-thick cladding-type layer of
ultra-pure copper on the cathode surface, which is expected to reduce
the background in the said energy range by approximately 55\%. This
procedure uses established techniques~\cite{Hoppe2008, HOPPE2007486},
and has been implemented by earlier
experiments~\cite{Abgrall:2013rze}.

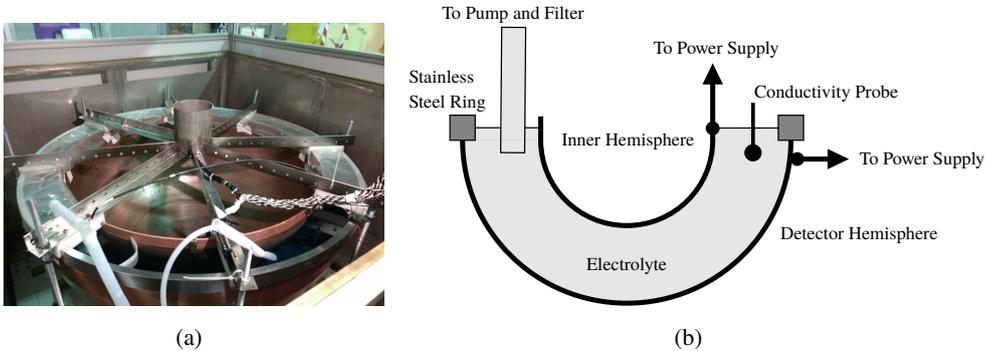
\begin{figure}[h]
\centering
\subfigure[\label{subfig:hemi1Installed}]{\includegraphics[width=0.33\textwidth]{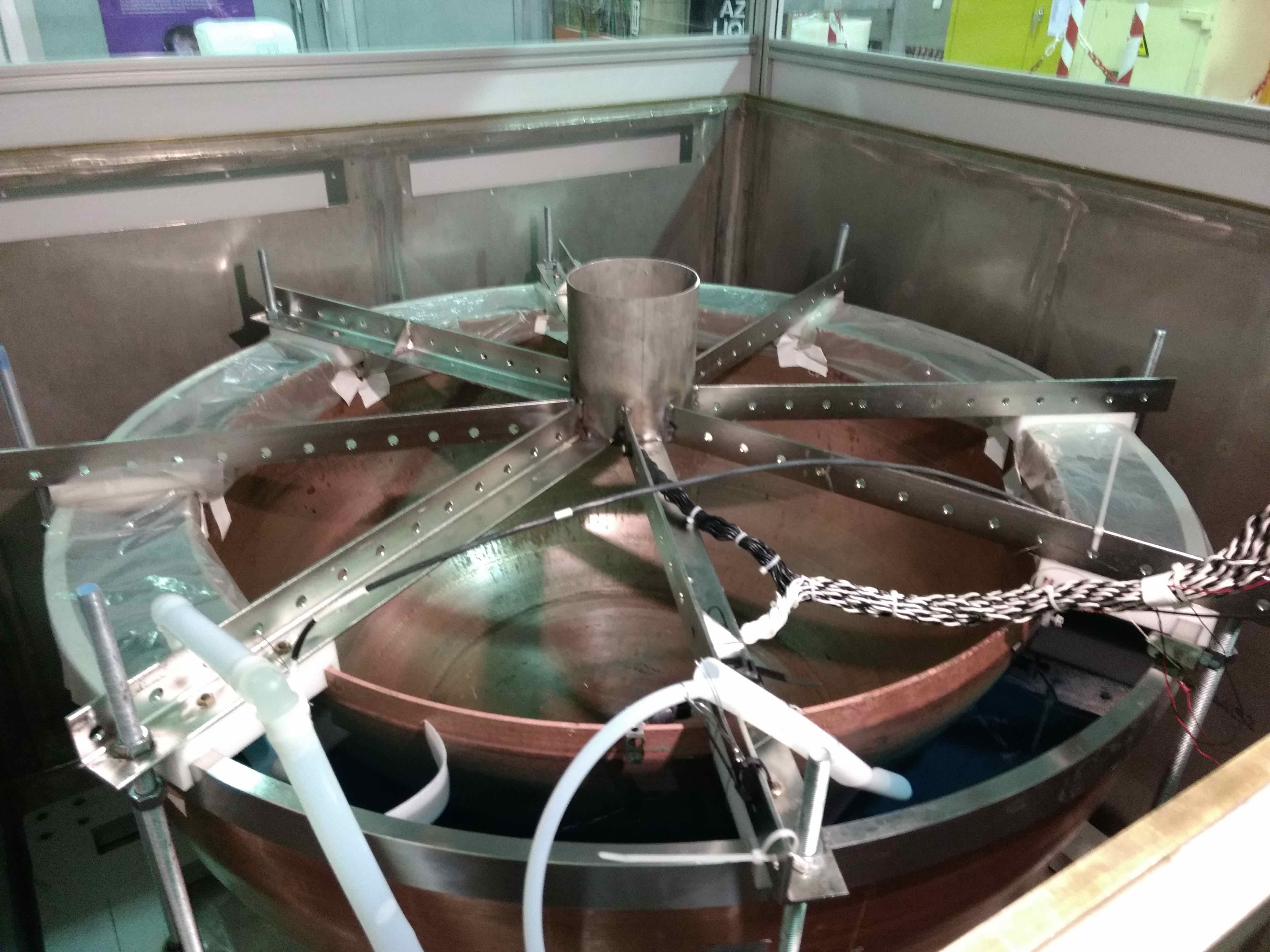}}
\subfigure[\label{subfig:hemisphereElectrolysis}]{ \resizebox{.52\textwidth}{!}{\tikzset{every picture/.style={line width=0.75pt}} %set default line width to 0.75pt        

\begin{tikzpicture}[x=0.75pt,y=0.75pt,yscale=-1,xscale=1]

%Shape: Rectangle [id:dp2581054489506507] 
\draw  [draw opacity=0][fill={rgb, 255:red, 0; green, 0; blue, 0 }  ,fill opacity=0.1 ] (279.04,140.22) -- (340.95,140.22) -- (340.95,150.22) -- (279.04,150.22) -- cycle ;
%Shape: Rectangle [id:dp674790359937226] 
\draw  [draw opacity=0][fill={rgb, 255:red, 0; green, 0; blue, 0 }  ,fill opacity=0.1 ] (80,140) -- (141.91,140) -- (141.91,150) -- (80,150) -- cycle ;
%Shape: Block Arc [id:dp607640888630838] 
\draw  [draw opacity=0][fill={rgb, 255:red, 0; green, 0; blue, 0 }  ,fill opacity=0.1 ] (340.95,150.22) .. controls (340.4,221.9) and (281.57,279.85) .. (209.53,279.66) .. controls (137.58,279.47) and (79.66,221.38) .. (80,149.83) -- (143.83,149.83) .. controls (143.66,186.2) and (173.25,215.74) .. (209.99,215.83) .. controls (246.79,215.93) and (276.84,186.47) .. (277.12,150.03) -- cycle ;
%Shape: Arc [id:dp7041995478955307] 
\draw  [draw opacity=0][line width=3]  (340,149.07) .. controls (340,149.38) and (340,149.69) .. (340,150) .. controls (340,221.8) and (281.8,280) .. (210,280) .. controls (138.2,280) and (80,221.8) .. (80,150) -- (210,150) -- cycle ; \draw  [line width=3]  (340,149.07) .. controls (340,149.38) and (340,149.69) .. (340,150) .. controls (340,221.8) and (281.8,280) .. (210,280) .. controls (138.2,280) and (80,221.8) .. (80,150) ;
%Shape: Arc [id:dp16917229565688707] 
\draw  [draw opacity=0][line width=3]  (278.08,149.51) .. controls (278.08,149.68) and (278.09,149.84) .. (278.09,150) .. controls (278.09,187.6) and (247.6,218.09) .. (210,218.09) .. controls (172.4,218.09) and (141.91,187.6) .. (141.91,150) -- (210,150) -- cycle ; \draw  [line width=3]  (278.08,149.51) .. controls (278.08,149.68) and (278.09,149.84) .. (278.09,150) .. controls (278.09,187.6) and (247.6,218.09) .. (210,218.09) .. controls (172.4,218.09) and (141.91,187.6) .. (141.91,150) ;
%Shape: Rectangle [id:dp9292228741673612] 
\draw  [fill={rgb, 255:red, 131; green, 131; blue, 131 }  ,fill opacity=1 ] (70,130) -- (90,130) -- (90,149.99) -- (70,149.99) -- cycle ;
%Shape: Rectangle [id:dp03655308239321231] 
\draw  [fill={rgb, 255:red, 131; green, 131; blue, 131 }  ,fill opacity=1 ] (330,130.01) -- (350,130.01) -- (350,150) -- (330,150) -- cycle ;
%Straight Lines [id:da1693957483292352] 
\draw [line width=3]    (141.91,130.95) -- (141.91,151) ;

%Straight Lines [id:da4098083854986003] 
\draw [line width=3]    (278.08,130.46) -- (278.08,150.51) ;

%Straight Lines [id:da022709047561390916] 
\draw    (90,140) -- (110,140) ;

%Straight Lines [id:da975451666706874] 
\draw    (279.04,140.22) -- (330.95,140.21) ;

%Straight Lines [id:da858503598888859] 
\draw [line width=3]    (278.08,93.46) -- (278.08,131.46) ;

\draw [shift={(278.08,88.46)}, rotate = 90] [fill={rgb, 255:red, 0; green, 0; blue, 0 }  ][line width=3]  [draw opacity=0] (16.97,-8.15) -- (0,0) -- (16.97,8.15) -- cycle    ;
%Shape: Circle [id:dp12190545045008161] 
\draw  [fill={rgb, 255:red, 0; green, 0; blue, 0 }  ,fill opacity=1 ] (273.08,140.49) .. controls (273.08,137.73) and (275.32,135.49) .. (278.08,135.49) .. controls (280.84,135.49) and (283.08,137.73) .. (283.08,140.49) .. controls (283.08,143.25) and (280.84,145.49) .. (278.08,145.49) .. controls (275.32,145.49) and (273.08,143.25) .. (273.08,140.49) -- cycle ;
%Shape: Circle [id:dp463009341351563] 
\draw  [fill={rgb, 255:red, 0; green, 0; blue, 0 }  ,fill opacity=1 ] (340,165) .. controls (340,162.24) and (342.24,160) .. (345,160) .. controls (347.76,160) and (350,162.24) .. (350,165) .. controls (350,167.76) and (347.76,170) .. (345,170) .. controls (342.24,170) and (340,167.76) .. (340,165) -- cycle ;
%Straight Lines [id:da7009929231883685] 
\draw [line width=3]    (380.94,165) -- (345,165) ;

\draw [shift={(385.94,165)}, rotate = 180] [fill={rgb, 255:red, 0; green, 0; blue, 0 }  ][line width=3]  [draw opacity=0] (16.97,-8.15) -- (0,0) -- (16.97,8.15) -- cycle    ;
%Shape: Rectangle [id:dp6904788980096523] 
\draw   (130,60) -- (129.16,159.99) -- (110.17,159.83) -- (111.01,59.84) -- cycle ;
%Straight Lines [id:da9016565072872567] 
\draw    (129.09,140) -- (142.91,140) ;

%Shape: Rectangle [id:dp3755913705863083] 
\draw  [draw opacity=0][fill={rgb, 255:red, 0; green, 0; blue, 0 }  ,fill opacity=0.1 ] (111.01,59.84) -- (130,59.84) -- (130,140) -- (111.01,140) -- cycle ;
%Straight Lines [id:da8483124927291823] 
\draw [line width=2.25]    (310,120) -- (310,160) ;
\draw [shift={(310,160)}, rotate = 90] [color={rgb, 255:red, 0; green, 0; blue, 0 }  ][fill={rgb, 255:red, 0; green, 0; blue, 0 }  ][line width=2.25]      (0, 0) circle [x radius= 5.36, y radius= 5.36]   ;

% Text Node
\draw (394,226) node  [align=left] {Detector Hemisphere};
% Text Node
\draw (211,151) node  [align=left] {Inner Hemisphere};
% Text Node
\draw (210,250) node  [align=left] {Electrolyte};
% Text Node
\draw (68,111) node  [align=left] {Stainless \\Steel Ring};
% Text Node
\draw (281,78) node  [align=left] {To Power Supply};
% Text Node
\draw (444,166) node  [align=left] {To Power Supply};
% Text Node
\draw (120,50) node  [align=left] {To Pump and Filter};
% Text Node
\draw (368,113) node  [align=left] {Conductivity Probe};

\end{tikzpicture}}}
  \caption{\subref{subfig:hemi1Installed} Electroplating set-up showing the detector hemisphere, anode, support structures and fixtures. \subref{subfig:hemisphereElectrolysis} schematic of the set-up. A pump is installed to provide mechanical mixing, while the attached filter removes particulates greater than $1\;\si{\micro\meter}$ in size from
the electrolyte.\label{fig:hemisphereElectrolysisFigs}}
\end{figure}

The electroplating was performed in LSM and the set-up is shown in figure~\ref{fig:hemisphereElectrolysisFigs}. 
A smaller hemisphere of copper was
prepared to act as the anode for electroplating. This smaller
hemisphere was suspended concentrically inside the detector
hemisphere, separated by an electrolyte of de-ionised water and
sulphuric acid. During electroplating, a voltage between the electrodes 
induces a current through the electrolyte solution. This
facilitates reduction reactions at the cathode which result in the
deposition of ions from the electrolyte on the cathode surface.

The hemispheres were first cleaned and sanded, to remove raised
portions of copper. The surface was then chemically etched using an
acidified hyrdrogen peroxide solution~\cite{HOPPE2007486}.
Subsequently, the surface was electropolished, the opposite process of
electroplating, removing $\left(21.2\pm0.1\right)\;\si{\micro\meter}$ and
$\left(28.2\pm0.1\right)\;\si{\micro\meter}$ from each of the detector
hemispheres, respectively, to ensure its smoothness, to expose the
underlying crystal structure, and to introduce copper ions into the electrolyte.

The electroplating continued for approximately 15 days for each
hemisphere using a pulse reverse current
technique~\cite{CHANDRASEKAR20083313}. The potential difference used
between the anode and the cathode for electroplating was
$0.3\;\si{\volt}$, the established value for electroplating pure
copper. From the recorded current, and by assuming uniform deposition
of copper, it was estimated that $(502.1\pm0.2)\;\si{\micro\meter}$
and $(539.5\pm0.2)\;\si{\micro\meter}$ of copper were plated onto the
surface of the two hemispheres, respectively. Finally, the surface was
rinsed with water and passivated with a $1\%$ citric acid
solution~\cite{HOPPE2007486}. The achieved plating rate corresponds to
approximately $1.3\;\si{\centi\meter\per\YEAR}$, demonstrating the
possibility to electroform a complete sphere underground.

\section{Gas purification}

\begin{wrapfigure}{r}{0.4\textwidth}
\centering
\includegraphics[width=0.39\textwidth]{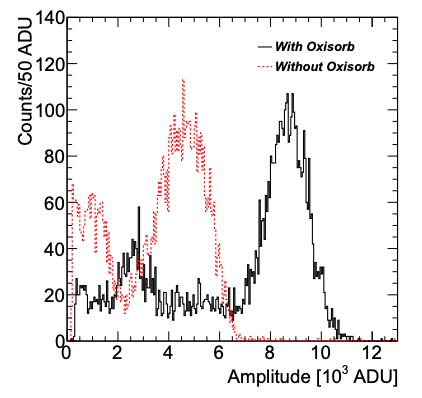}
\caption{\label{filter} Signal comparison of 5.9 keV X-rays in a
  spherical proportional counter filled with 600 mbar of He:CH$_{4}$
  (90\%:10\%) gas with and without filtering \cite{Knights:2019tmx}.}
\end{wrapfigure}
The quality of the gas mixture is of paramount importance for the
performance of the experiment. Electronegative gas contaminants lead
to electron attachment, and, thus, signal reduction and deterioration
of energy resolution and background discrimination. Such effects are
particularly important in regions of low electric field strength. Gas
filtering using Messer Oxisorb or Saes MicroTorr Purifier was
introduced to ensure that oxygen and water induced effects are
minimised. Figure~\ref{filter} shows the pulse amplitude for 5.9~keV
X-rays measured with a spherical proportional counter filled with
filtered and unfiltered gas. Filtering improved the measured resolution
(\textsigma/E), from $\left(21.3\pm0.7\right)\%$ to $\left(9.4\pm 0.3\right)\%$. However, it
was found that the filtering process introduces non-negligible amounts
of $^{222}$Rn. Such behaviour has been previously
reported~\cite{Calvo:2016hve, Novella:2018ewv,Avenier:2001ym}, and the
introduction of a carbon filter to remove the emanated $^{222}$Rn is
investigated.

\section{Detector calibration and monitoring}
\begin{figure}[h]
  \centering
\subfigure[\label{laser}]{\includegraphics[width=0.73\textwidth]{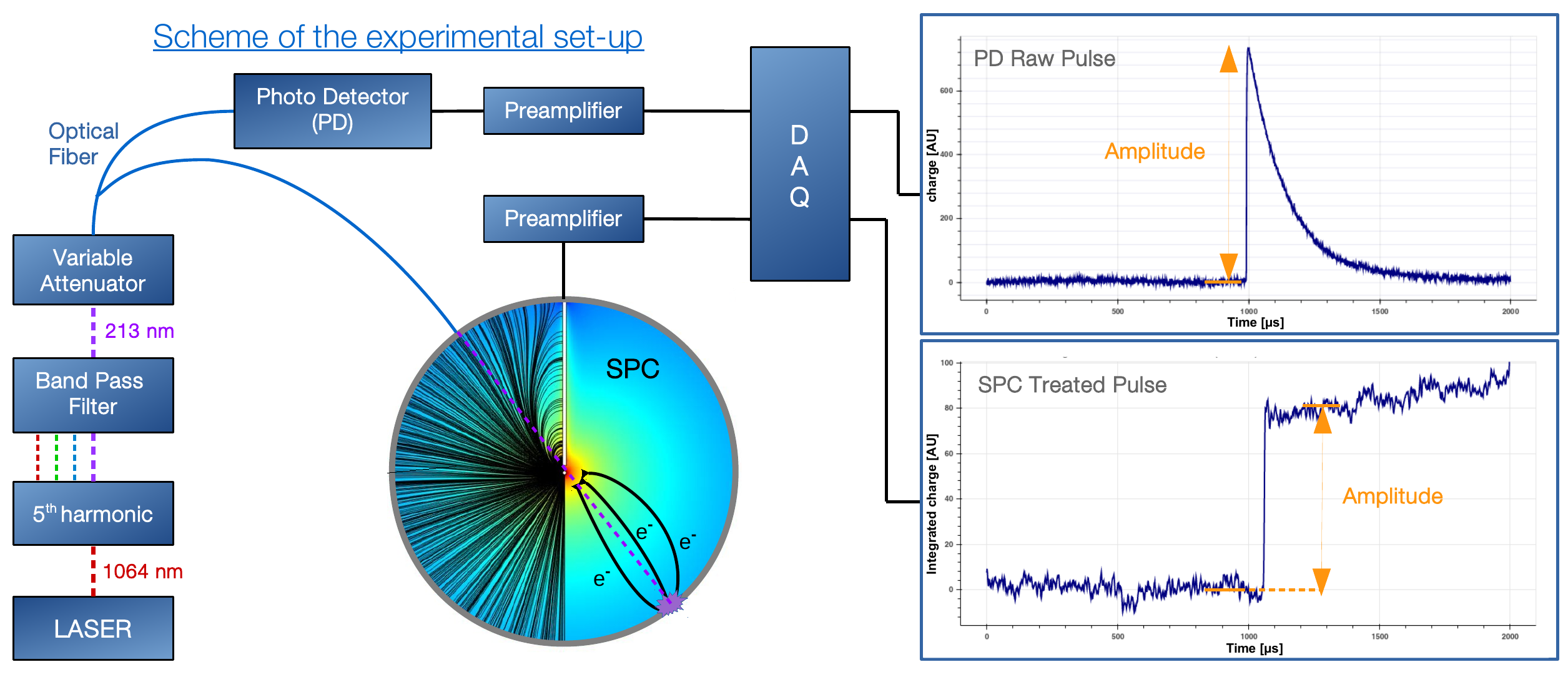}}
\subfigure[\label{laser2}]{\includegraphics[width=0.26\textwidth]{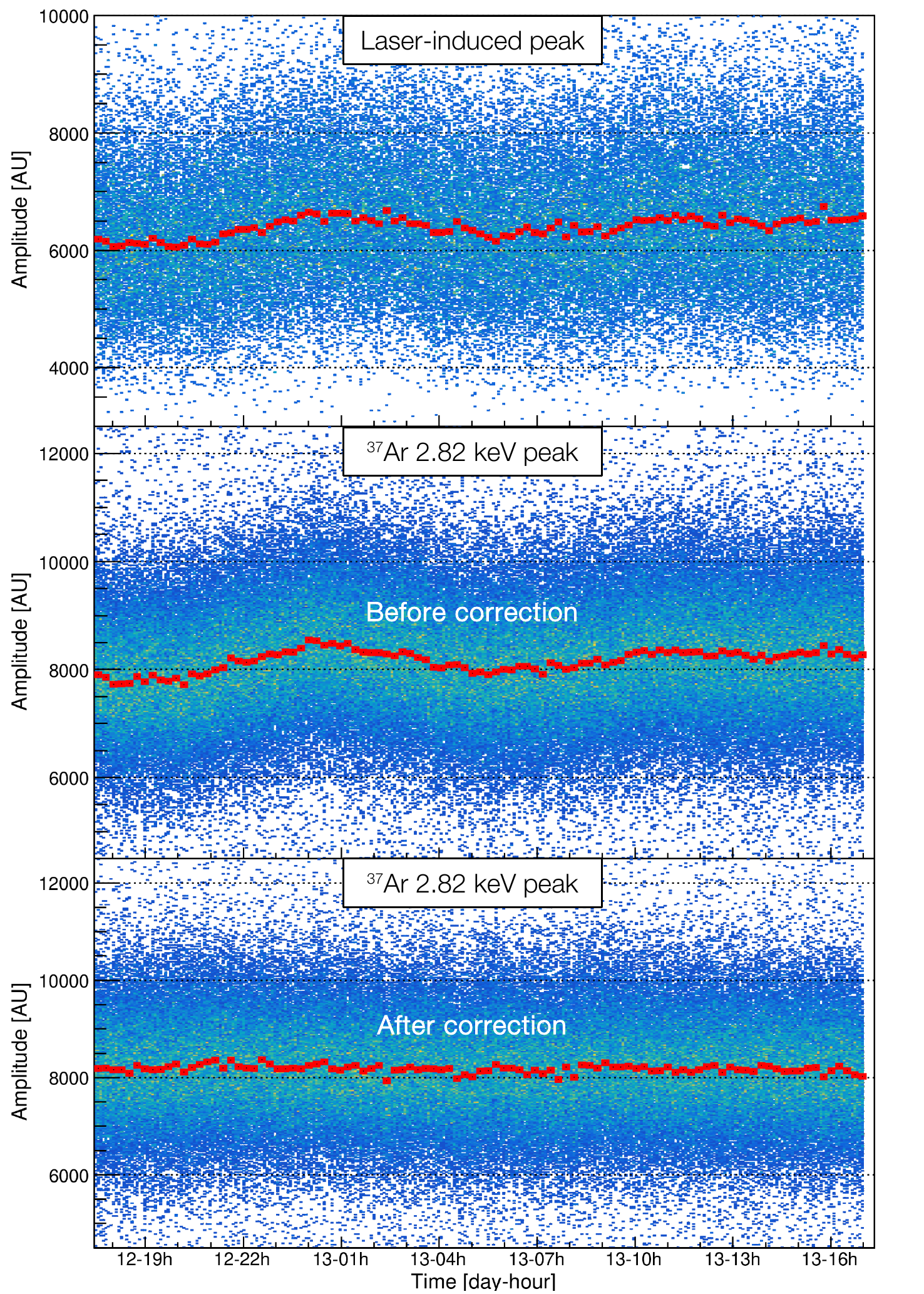}}
\caption{\subref{laser} Schematic of the laser-based calibration system. \subref{laser2} Gain stability monitoring in a spherical proportional counter using a UV laser. The top panel shows the distribution of
pulse amplitude as a function time for laser-induced events. The middle (bottom) panel shows the distribution of $^{37}$Ar 2822~eV events before (after) correcting for gain variations using the amplitude of laser-induced
events. Red markers indicate the centre of a Gaussian fit to amplitude spectra for time slices of 15~min. Both figures from ref.~\cite{Arnaud:2019nyp}.}
\end{figure}

The performance of the detector as a function of time, as well as
detailed studies of the detector response at the single electron level,
are enabled by the use of a laser-based calibration
system~\cite{Arnaud:2019nyp}. The experimental setup, which
incorporates a monochromatic UV laser beam with variable intensity, is
shown in figure~\ref{laser}. The trigger for data acquisition is
provided by the laser signal in a photo detector, allowing for precise
measurements of electron transport parameters, e.g. drift time,
diffusion coefficients, and electron avalanche gain, as well as
trigger efficiency measurements. These studies are complement with an $^{37}$Ar gaseous calibration source for measurements of the gas
W-value and Fano factor. The calibration system can be used in
parallel with data-taking for physics to monitor the detector
response, as shown in figure~\ref{laser2}.

\section{Summary}
The spherical proportional counter is a novel gaseous detector
offering significant advantages in the search for light dark matter
candidates in the range between 0.1 and 10~GeV. The first physics
results of the NEWS-G collaboration were obtained using the SEDINE
detector, a 60~cm in diameter spherical proportional counter,
operating at LSM. These results demonstrate the potential of spherical
proportional counters in the search of low-mass DM candidates.
A new 140~cm in diameter detector, was constructed and commissioned at
LSM during summer 2019, and is currently being installed in
SNOLAB. Several improvements, based on the experience from SEDINE,
have been incorporated in the design. These include low-activity
copper, electroplating a layer of copper onto the inner surface of the
detector, and a new, compact shielding. These improvements, together
with the developments in sensor design, are expected to lead to
substantial improvements in the sensitivity for light DM candidates.

\section*{Acknowledgments}
This work is supported by the Royal Society International Exchanges
and the IPPP Associateship schemes, and by UKRI-STFC through the
University of Birmingham Particle Physics Consolidated Grant.

\bibliographystyle{JHEP}
\bibliography{H4lpaper}

\end{document}